\documentclass[superscriptaddress,reprint,aps,prl]{revtex4-1}
\usepackage{amssymb, amsmath}
\usepackage{graphicx}% Include figure files
\usepackage{dcolumn}% Align table columns on decimal point
\usepackage{bm}% bold math
\usepackage{float}
\usepackage{xcolor}
\usepackage[normalem]{ulem}

\begin{document}

\title{ $\mu$SR study of spin freezing and persistent spin dynamics in NaCaNi$_2$F$_7$}

\author{Y. Cai}
\author{M.N. Wilson}
\author{A.M. Hallas}
\address{Department of Physics and Astronomy, McMaster University, Hamilton, Ontario L8S 4M1, Canada}
 
\author{L. Liu}
\author{B. A. Frandsen}
\address{Department of Physics, Columbia University, New York, New York 10027, USA}
 
 \author{S. R. Dunsiger}
 \address{Center for Emergent Materials, The Ohio State University, Columbus, OH 43210, USA}
 
 \author{J. W. Krizan}
 \author{R. J. Cava}
 \address{Department of Chemistry, Princeton University, Princeton, NJ, 08544, USA}
 
\author{Y. J. Uemura}
\address{Department of Physics, Columbia University, New York, New York 10027, USA}
 
 \author{O.~Rubel}
\address{Department of Materials Science and Engineering, McMaster University, Hamilton, Ontario L8S 4M1, Canada}
\author{G.M. Luke}
\address{Department of Physics and Astronomy, McMaster University, Hamilton, Ontario L8S 4M1, Canada}
\address{Canadian Institute for Advanced Research, Toronto, Ontario M5G 1M1, Canada}

\begin{abstract}
A new pyrochlore compound, NaCaNi$_2$F$_7$, was recently synthesized and has a single magnetic site with spin-1 Ni$^{2+}$. We present zero field (ZF) and longitudinal field (LF) muon spin rotation ($\mu$SR) measurements on this pyrochlore. Density functional theory (DFT) calculations show that the most likely muon site is located between two fluorine ions, but off-centre. A characteristic F-$\mu$-F muon spin polarization function is observed at high temperatures where Ni spin fluctuations are sufficiently rapid. The Ni$^{2+}$ spins undergo spin freezing into a disordered ground state below 4~K, with a characteristic internal field strength of 140~G. Persistent Ni spin dynamics are present to our lowest temperatures (75~mK), a feature characteristic of many geometrically frustrated magnetic systems.
\end{abstract}

\maketitle

A rich variety of magnetic ground states have been realized in the \textit{A$_2$B$_2$X$_7$} family of geometrically frustrated pyrochlores, including spin glasses, spin liquids, spin ices, as well as long-range-ordered states \cite{Jason S. Gardner, hallas_arxiv}. Both the \textit{A} and \textit{B} sites form corner-sharing tetrahedral networks that can host many different ions which allows broad tuning of the physical interactions. For example, in titanium \textit{B}-site pyrochlores, placing terbium on the \textit{A}-site gives an enigmatic spin liquid state, holmium and dysprosium give a classical spin ice state, while gadolinium gives a unique partially ordered state \cite{Gardner, harris_PRL, Ramirez}. The competing energy scales in these systems (spin anisotropy, exchange, and dipolar interactions) as well as sensitivity to dilute disorder have emerged as extremely important in selecting the ground states of these materials \cite{Stewart, Jason S. Gardner}. 

One interesting state that occurs in some pyrochlores is the spin glass. Although classical spin glass materials have been well studied \cite{Binder1986}, open questions remain about how such a ground state might emerge in frustrated pyrochlores \cite{Jason S. Gardner}. Classical spin glass states typically occur in dilute magnetic alloys, where the magnetic state arises from the long-range oscillatory RKKY interaction coupled with a distribution of spin$-$spin separations \cite{M. A. Ruderman, K. Yosida}. A similar state could occur in pyrochlores arising from local disorder of the spin direction, yielding an effective dilute magnetic system. Other theoretical studies show the possibility that bond disorder in pyrochlores could result in a spin glass state \cite{T. E. Saunders, A. Andreanov, H. Shinaoka}, which was observed in nominally disorder free Y$_2$Mo$_2$O$_7$ \cite{Greedan, Silverstein, Thygesen}.

Recently, fluoride pyrochlores in the \textit{A$_2$B$_2$F$_7$} family have been synthesized, expanding the diversity of pyrochlore systems \cite{Krizan2014, Krizan2015, Sanders2017, R. J. Cava}. In these materials, the $A$ site is presumably randomly occupied by two different elements from Group 1 and 2 on the periodic table, and the $B$ site is occupied by a magnetic transition metal. The chemical disorder on the $A$ site leads to bond disorder around the magnetic site that is believed to substantially influence their low temperature properties. This new family of magnetic pyrochlores provides a platform for the study of frustrated magnetism where the strong electronic correlations of the transition metals, along with bond disorder caused by the random occupation of the $A$ site, may show unique magnetic properties. 

All six known members of this family show spin freezing transitions at temperatures below 4~K, with antiferromagnetic Curie-Weiss temperatures larger than 50~K, indicating substantial frustration. AC susceptibility measurements suggest that NaCaFe$_2$F$_7$, NaSrFe$_2$F$_7$, and NaSrMn$_2$F$_7$ each have a spin glass ground state below this spin freezing transition \cite{Sanders2017}. Neutron scattering shows short range XY-antiferromagnetism in NaCaCo$_2$F$_7$ and NaSrCo$_2$F$_7$, with substantial gapless fluctuations through different local XY states \cite{Ross2016, Ross2017}, and nuclear magnetic resonance (NMR) shows spin glass like freezing at 3.6~K with thermally activated spin fluctuation at low  temperature in NaCaCo$_2$F$_7$ \cite{Sarkar_NMR}. The lack of long range order in these Co pyrochlores is suggested to come from ice-like correlated disorder between the Na and Ca or Sr on the $A$ site \cite{Ross2016}. The spin-$1$ Ni$^{2+}$ pyrochlore, NaCaNi$_2$F$_7$, has also been synthesized and shown to have spin freezing at 3.6~K with antiferromagnetic exchange interactions and a corresponding frustration index of $f\sim36$ \cite{R. J. Cava}. The magnetic entropy integrates to the value for a Ising system, R ln(2), which is less than expected for an S = 1 system, indicating the presence of significant residual entropy. Surprisingly, this Ni pyrochlore releases more entropy than in NaCaCo$_2$F$_7$ \cite{R. J. Cava}. A bond-disorder induced spin glass state is a likely candidate for the ground state of this system, although no microscopic measurements of the magnetism have been reported. Local magnetic probes that are sensitive to the real space field distribution and to spin fluctuations would be useful to better understand this magnetic state.

In this paper, we present muon spin rotation and relaxation ($\mu$SR) measurements on high quality single crystals of NaCaNi$_2$F$_7$ \cite{R. J. Cava}. $\mu$SR can probe both static magnetic order and dynamic spin fluctuations, which has been useful in studying conventional spin glass systems such as Cu$_{x}$Mn and AuFe$_{x}$ \cite{Uemura3, Uemura4}. In NaCaNi$_2$F$_7$, we observe the slowing down of the Ni spins when cooling down to 4 K, with a spectra absent of coherent oscillation at base temperature, and dynamic spin fluctuations that persist even at 75~mK. Detailed analysis of the relaxation function shows that these fluctuations are consistent with the persistent spin dynamics commonly seen in other frustrated systems \cite{A. M. Hallas, Uemura2, A. Yaouanc, J. Lago, 2Dunsiger}. We also performed density functional theory (DFT) \cite{Kohn_PR_140_1965} electronic structure calculations to determine the muon stopping sites of NaCaNi$_2$F$_7$, suggesting an off centered muon position between two adjacent fluorine ions, whose positions are distorted (drawn together) by the presence of the muon.

The NaCaNi$_2$F$_7$  sample was grown in an optical floating zone furnace based on a modified Bridgman-Stockbarger method as described previously \cite{R. J. Cava}. We performed $\mu$SR measurements on the M15 and M20 beam-lines at the TRIUMF laboratory in Vancouver, Canada, with dilution refrigerator and He-4 gas flow cryostat setups. One large crystal of NaCaNi$_2$F$_7$ was mounted on the M20 beam-line with the {(}110{)} direction parallel to the incoming beam in the He-4 cryostat low background apparatus . This sample was subsequently sliced into $\sim$1mm thick discs along the same {(}110{)} direction and then mounted onto an Ag plate and covered in thin Ag foil for the measurements in the dilution refrigerator. We performed measurements in both zero applied field and with a magnetic field applied along the incident muon spin direction. All the $\mu$SR data were fit using the open source $\mu$SRfit software package \cite{A. Suter}.

In $\mu$SR measurements, spin polarized muons are implanted into a sample one at a time where they thermalize rapidly in the material while maintaining their polarization. These thermalized muons find a minimum electrostatic potential site where they come to rest and their spins precess in the local magnetic field until they decay with a average lifetime $\tau_{\mu}=2.2~{\mu}s$, emitting a positron preferentially in the direction of the muon spin. In materials containing fluorine, muons generally thermalize close to these highly electronegative fluorine ions \cite{J. H. Brewer}. At high temperature, well above any magnetic transition, where the electronic fluctuations are exchange narrowed and too rapid to contribute to the muon spin relaxation signal, the signal in such compounds typically shows oscillations attributed to the  F-$\mu$-F magnetic dipole interaction. This is caused by the dipole-dipole interactions between the muon and the surrounding fluorine nuclei (usually two fluorines). By modelling the simplest configuration of muons lying at the mid-point between two fluorine ions, the muon spin relaxation function $G_{F\mu{F}}$$(t)$ can be obtained by averaging over all different orientations of fluorine nuclear dipole moments \cite{J. H. Brewer, D. R. Noakes1, J. Rodriguez}. In a cubic system, this relaxation function is given by,
\begin{equation}
P(t)=A_{total} (f e^{-(\lambda_{1} t)^{\beta}} G_{F\mu{F}}(t) + (1-f) e^{-\lambda_{2} t})
\label{eq}
\end{equation}
with 
\begin{multline}
G_{F\mu{F}}= 1/6 ( 3 + \cos(\sqrt{3}\omega_{d}t) +
 \\
(1-1/\sqrt{3})\cos[((3-\sqrt{3})/2)\omega_{d}t] + 
\\
(1+1/\sqrt{3}) \cos[((3+\sqrt{3})/2)\omega_{d}t] ).
\label{eq2}
\end{multline}
Here, $\omega_d$ is the muon-fluorine dipole frequency, $\beta$ is a phenomenological stretching parameter, $\lambda_1$ is the relaxation rate of the F-$\mu$-F signal that comes mainly from the effects of further near neighbours, and $\lambda_2$ is due to background that is assumed to be a simple exponential. The fraction $f$ represents the F-$\mu$-F signal fraction and $1-f$ represents the background fraction not associated with the F-$\mu$-F signal. All the asymmetries are scaled by $A_{total}$ in the following figures.

\begin{figure}[h]
\setcounter{bottomnumber}{6}
\includegraphics[width=\columnwidth]{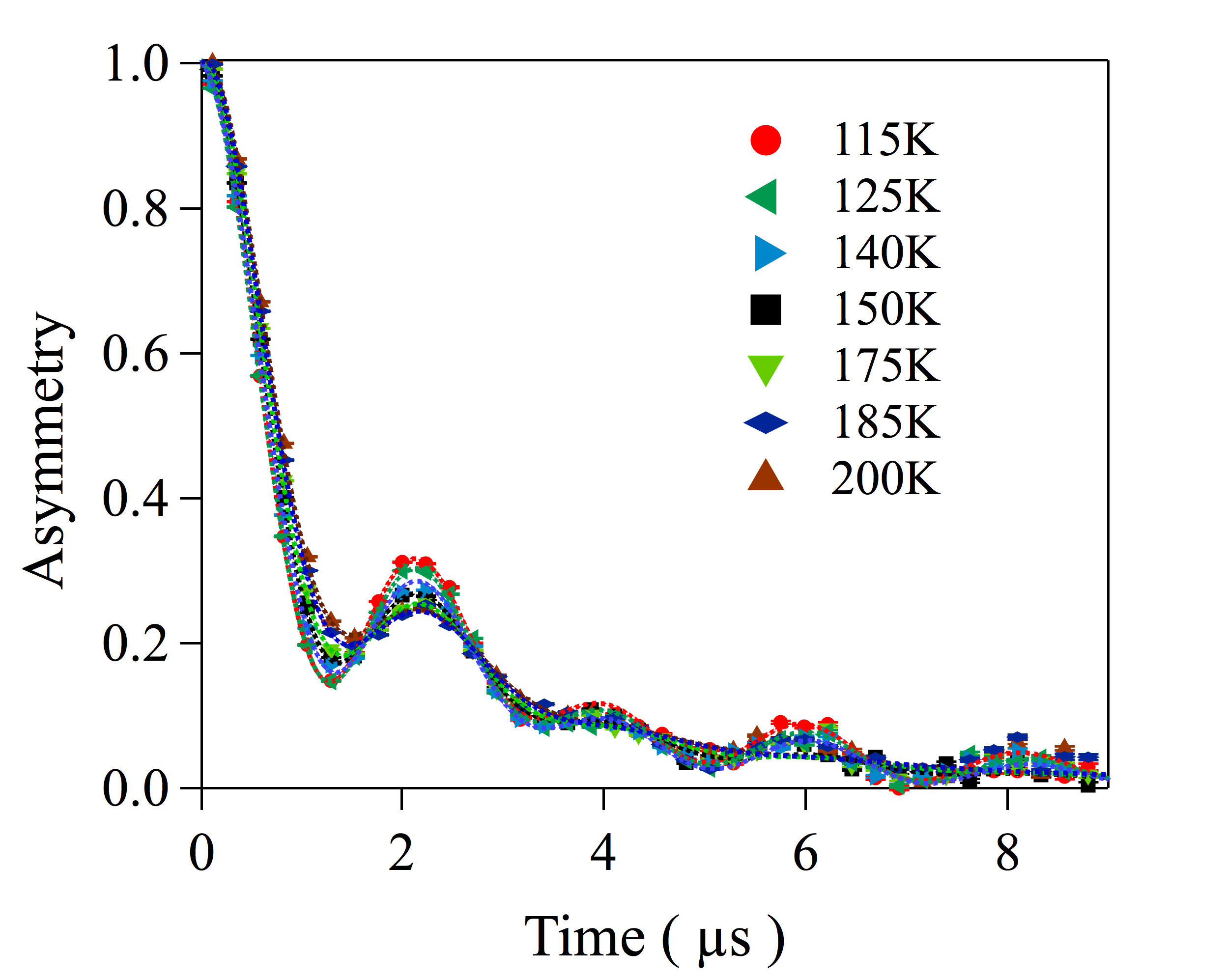}
\caption{ZF-$\mu$SR time spectra measured in NaCaNi$_2$F$_7$ above 100K. Oscillations due to the F-$\mu$-F signal are well defined.}
\label{Fig: FuF2}
\end{figure}

\begin{figure}[h]
\setcounter{bottomnumber}{6}
\includegraphics[width=\columnwidth]{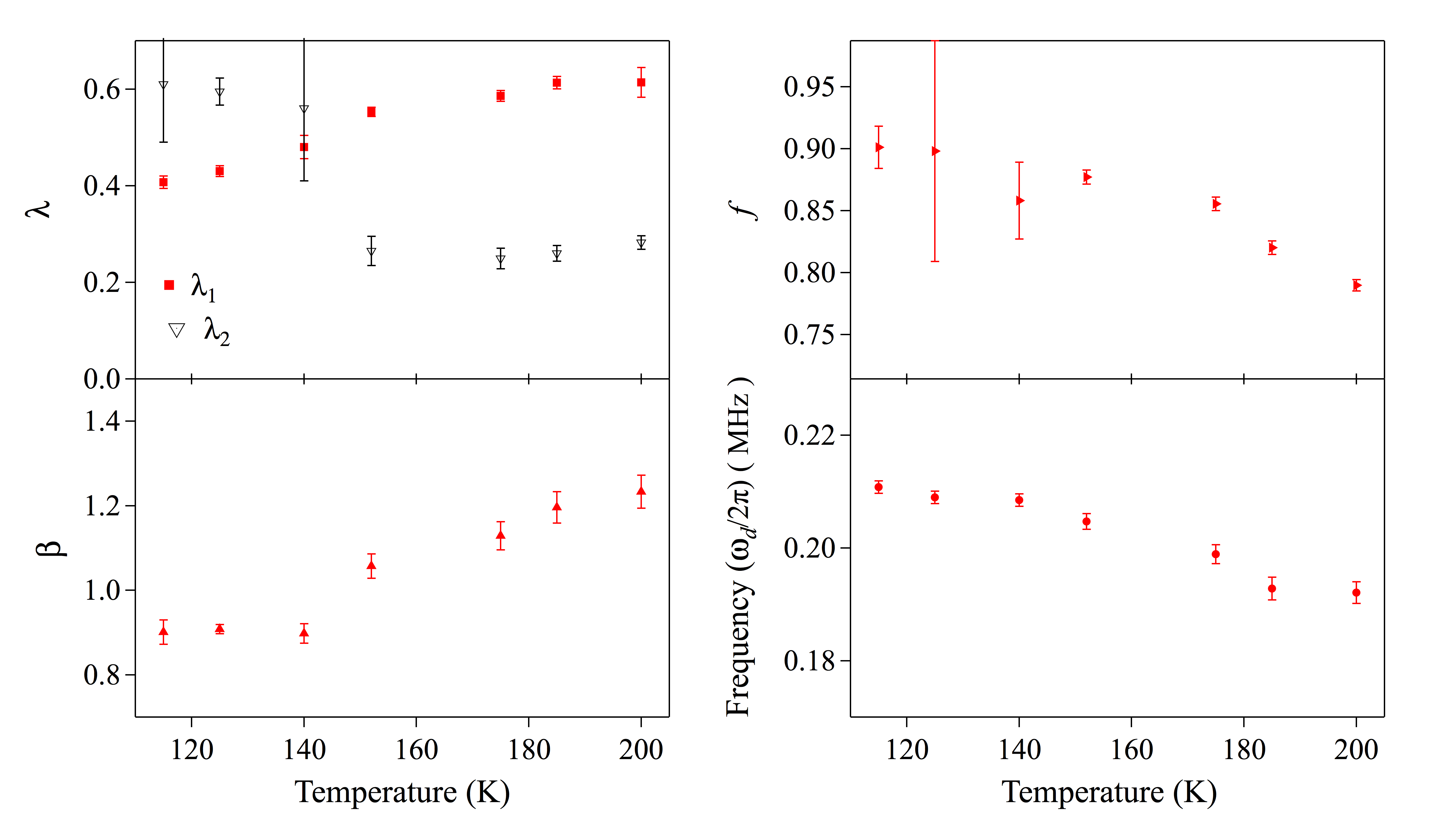}
\caption{Temperature dependence of the fitting parameters of Eq 1 used to fit the high temperature NaCaNi$_2$F$_7$ ZF $\mu$SR data.}
\label{Fig: fitparameters}
\end{figure}

We show data for our NaCaNi$_2$F$_7$ sample between 100 and 200~K in Figure~\ref{Fig: FuF2}, which is well above the spin freezing temperature of Ni moments. This data shows clear F-$\mu$-F oscillations that we fit using Eq \ref{eq} to extract the fitting parameters shown in Figure~\ref{Fig: fitparameters}. Since the frequency is mainly determined by the peak position and not that sensitive to the relaxation rate in the spectra, we allowed the other parameters to freely vary to obtain the best possible fit. The fit value of $\beta \approx 1$ corresponds to the typical exponential behaviour, and the fraction $f$ is close to 90\%. The fitted frequency is slightly dependent on temperature as shown in Figure~\ref{Fig: fitparameters} and has an average value of $\omega_d$/2$\pi$ = 0.201(8)~MHz over the 120~K to 200~K temperature range. The frequency temperature dependence may be due to a subtle change in the muon site, or more likely a trading-off effect in the fitting parameters. This value corresponds to a F-$\mu$-F bond length of 2$r$ = 2.420(4)~\AA\ \cite{J. H. Brewer}. Based on the crystal structure (Ref \cite{R. J. Cava}), the flattened octahedra of NiF$_6$$^{4-}$ could result in three different F-F equilibrium bonds lengths : 2.656~\AA\, 2.828~\AA\ and 2.991~{\AA} . Therefore, our data indicates that there is significant local distortion to the fluorine positions caused by the muon. This distortion effect is frequently seen, and is caused by the strength of the electrostatic attraction between the muons and fluorine ions \cite{J. H. Brewer, D. R. Noakes2, G. M. Luke}.

 \begin{figure*}%[h]
	\includegraphics[width=0.9\textwidth]{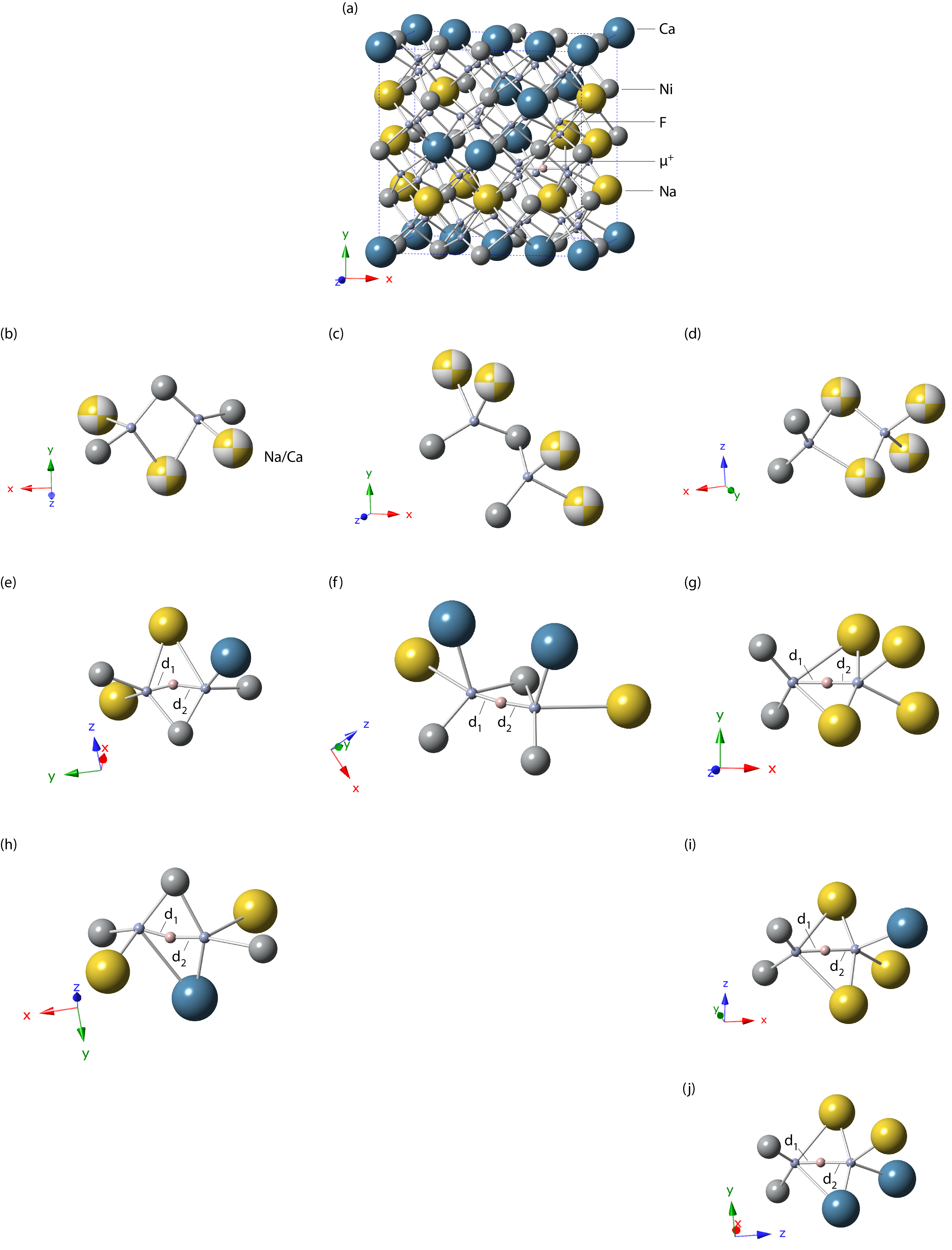}
	\caption{Possible $\mu^+$ implantation sites in NaCaNi$_2$F$_7$. (a) The crystal structure of an 88-atom supercell. (b-d) Generic local environments of Florine atoms that correspond to shortest (b), second shortest (c), and third shortest (d) F-F distances, which are referred as $\alpha$-Ni$_3$(Na,Ca)$_3$, $\beta$-Ni$_3$(Na,Ca)$_4$, and $\gamma$-Ni$_2$(Na,Ca)$_4$ sites, respectively. (e-j) The local structure of possible muon sites with a specific nearest-neighbor environment: (e) $\alpha$-Ni$_3$Na$_2$Ca, (f) $\beta$-Ni$_3$Na$_2$Ca$_2$, (g)  $\gamma$-Ni$_2$Na$_4$, (h) $\alpha$-Ni$_3$NaCa$_2$, (i) $\gamma$-Ni$_2$Na$_3$Ca, and (j) $\gamma$-Ni$_2$Na$_2$Ca$_2$. The muon is trapped nearly midway between the two fluorine atoms. The F-$\mu^+$ bond lengths $d_1$ and $d_2$ are generally not identical and are sensitive to the nearest-neighbor environment.}
	\label{Fig:sites}
\end{figure*}

To better understand the likely muon position, we performed electronic structure calculations. These calculations used the Perdew-Burke-Ernzerhof generalized gradient approximation \cite{Perdew_PRL_77_1996} (GGA) for the exchange-correlation functional with the Vienna \textit{ab initio} simulation program \cite{Kresse_PRB_47_1993,Kresse_PRB_54_1996,Kresse_CMS_6_1996} (VASP) and projector augmented-wave (PAW) potentials \cite{Kresse_PRB_59_1999,Blochl_PRB_50_1994}. The structure of NaCaNi$_2$F$_7$ was represented by a pseudo-cubic 88-atom supercell (Fig.~\ref{Fig:sites}a) that contains 8 formula units. Na and Ca atoms were randomly mixed on the same sub-lattice sites with a ratio of 1:1. The experimental value of the lattice parameter ($a=10.28$~{\AA}) was used \cite{R. J. Cava}.  The muon was modelled as a proton, i.e., H-atom with one electron subtracted from the model. This strategy was successfully applied in previous studies of muon implantation sites in florides \cite{Bernardini_PRB_87_2013,Moeller_PRB_87_2013}.  Atomic positions were optimized by minimizing the Hellmann-Feynman forces acting on atoms below 10~meV/{\AA}. The Brillouin zone was sampled using $2\times2\times2$ Monkhorst-Pack grid \cite{Monkhorst_PRB_13_1976}. The cutoff energy for a plane wave expansion was set at 500~eV, which is 25\% higher than the value recommended in the pseudo-potential file. The higher cutoff energy was essential for obtaining accurate structural parameters.

Dudarev \textit{et al.} \cite{Dudarev_PRB_57_1998} GGA+U framework was used to account for strong electronic correlations due to the localized d-orbital for Ni. The effective on-site Coulomb interaction (Hubbard $U$) between localized d-electrons was set at $U=6.4$~eV as suggested by Wang et al. \cite{Wang_PRB_73_2006}. Nickel atoms were initialized with randomly oriented magnetic moments of magnitude $1~\mu_\text{B}$, which increased up to $1.85~\mu_\text{B}$ when a converged electron density was obtained. Our calculation show that the moment is insensitive ($\pm0.01~\mu_\text{B}$) to the local environment of Ni-atoms and to the presence of muon.  The net magnetic moment of the supercell remained nearly zero.

\begin{table}
    \caption{Characteristics of the muon sites in NaCaNi$_2$F$_7$.}
    \label{Table:sites}
    \begin{ruledtabular}
        \begin{tabular}{l c c c c}
            Environment &  Total energy & \multicolumn{3}{c}{F-$\mu^+$ bond length ({\AA})} \\
            \cline{3-5}
            (see Fig.~\ref{Fig:sites}) & difference (eV) & $d_1$ & $d_2$ & $d_1$+$d_2$\\
            \hline
			$\gamma$-Ni$_2$Na$_4$ & 0 (reference) & 1.22 & 1.10 & 2.32\\
			$\gamma$-Ni$_2$Na$_3$Ca & 0.3 & 1.19 & 1.13 & 2.32\\
			$\alpha$-Ni$_3$Na$_2$Ca & 0.8 & 1.09 & 1.26 & 2.35\\
            $\gamma$-Ni$_2$Na$_2$Ca$_2$ & 1.0 & 1.07 & 1.27 & 2.34\\
			$\beta$-Ni$_3$Na$_2$Ca$_2$ & 1.1 & 1.15 & 1.16 & 2.31\\
			$\alpha$-Ni$_3$NaCa$_2$ & 1.3 & 1.11 & 1.18 & 2.29\\
        \end{tabular}
    \end{ruledtabular}
\end{table}

Six alternative muon implantation sites were explored as shown in Fig.~\ref{Fig:sites}(e-j). The configurations are labeled according to the second nearest-neighbor environment of the muon. Each site was initialized with the muon positioned equidistantly on the line between the fluorine atoms. The DFT total energies of alternative configurations were compared upon completion of the structural optimization. The $\gamma$-Ni$_2$Na$_4$ configuration (Fig.~\ref{Fig:sites}g) is energetically most favorable (see Table~\ref{Table:sites}) and its energy separation from other muon sites is much greater than the temperature $k_\text{B}T$ at which the experiment was performed. Unlike simple fluorine salts such as LiF where all fluorine atoms are equivalent, it is not the case for the pyrochlore fluorides. As a result, the muon shifts from the midway position between two fluorine atoms. In the case of the $\gamma$-Ni$_2$Na$_4$ configuration, the $\mu^+$-F bond length disparity is 0.12~{\AA}  (Table~\ref{Table:sites}). The highest disparity is predicted to occur for the Ca-rich $\gamma$-Ni$_2$(Na,Ca)$_4$ configuration (Fig.~\ref{Fig:sites}), and the disparity vanishes for the $\beta$-Ni$_3$Na$_2$Ca$_2$ site (Fig.~\ref{Fig:sites}f) as a result of the symmetry of the fluorine environment.
 
 \begin{figure}[h]
	\includegraphics[width=0.94\columnwidth]{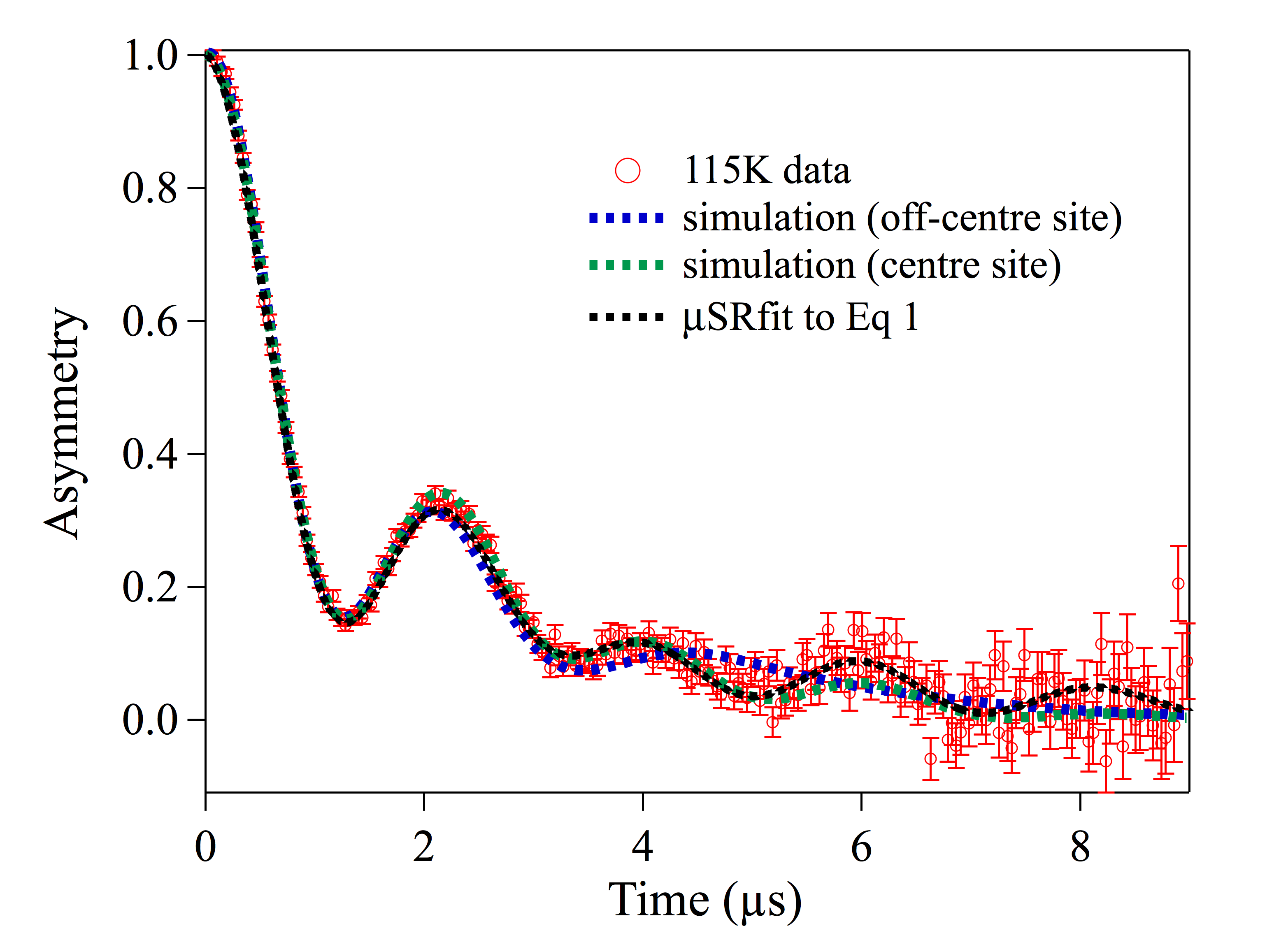}
	\caption{Different $\mu^+$ sites simulation in NaCaNi$_2$F$_7$ compared with the fitting result. Red points is the 115K data obtained in TRIUMF, the blue and green dashed lines are non-mid and mid muon site simulation results and the black dashed line is the fitting result from the $\mu$SRfit software which assumes a symmetric F-$\mu$-F site. }
	\label{Fig:cal-simu115K}
\end{figure}

We also performed simulations of the F-$\mu$-F spectra by a simple approach with two fluorines and one muon which gives a three spin quantum mechanical system. We confirmed the simulation by comparing with the classical theory calculation of Eq~\ref{eq2} assuming a muon site centred between two fluorine atoms, and find good agreement. Then, by offsetting the muon position, we found a possible asymmetric muon site simulation also captures most features of the spectra as shown in Fig.~\ref{Fig:cal-simu115K}. For this simulation, two fluorine positions were set to ($\pm$0.5,0,0)r$_{simu}$, and the off-centred simulation results in a muon site at (0.1,0.05,0.05)r$_{simu}$, with 2r$_{simu}$ = 2.4~{\AA}, which gives two different muon-fluorine bond of lengths of d$_1$=1.45$\pm$0.2~{\AA} and d$_2$=0.97$\pm$0.2~{\AA}. Comparing both the centred and off-centred simulations with the fitting result from $\mu$SRfit and the actual data, we find that the centre muon site configuration is not a unique final result for the observed F-$\mu$-F signal in this compound.

Apart from the expected muon site and its ZF-$\mu$SR relaxation function, we also calculated the approximate field magnitude at possible muon sites in the magnetically ordered state. Given the magnetic Ni moments structures from the DFT modelling above, we extended the 16-Ni cell in three different directions into a larger 16*3*3*3-Ni cell by assuming the periodic boundary conditions. The estimated dipolar field strengths are $\sim$ 2150~G and 3150~G for the lowest and next lowest energy muon position, corresponding to the $\gamma$-Ni$_2$Na$_4$ and $\gamma$-Ni$_2$Na$_3$Ca, respectively. In addition, for randomly oriented Ni moments, field strength at the lowest energy muon position was also calculated to be $\sim$ 1500~G. These estimated local fields are at least 15 times bigger than we observed experimentally, as will be shown later, which is evidence for strong dynamics in this compound.

\begin{figure}[h]
\setcounter{bottomnumber}{6}
\includegraphics[width=\columnwidth]{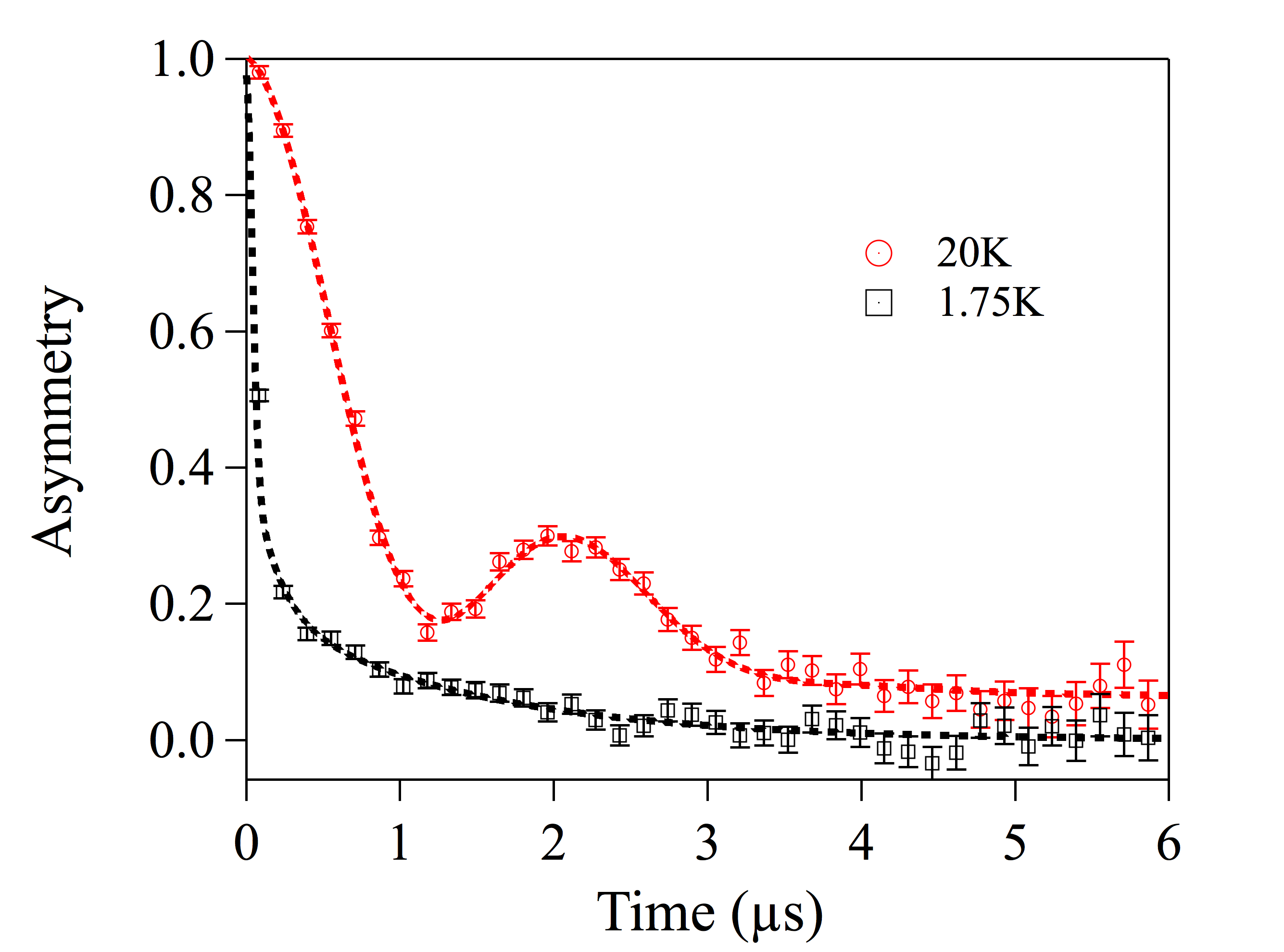}
\caption{Asymmetry spectra of the $\mu$SR data of NaCaNi$_2$F$_7$ at 20~K and 1.75~K in zero field. The oscillation at 20~K is due to the F-$\mu$-F signal, which is washed out at 1.75~K due to the frozen magnetic Ni moments. These spectra were measured on the M20 beamline}
\label{Fig: FuF}
\end{figure}

Typical low temperature $\mu$SR spectra of NaCaNi$_2$F$_7$ are shown in Fig.~\ref{Fig: FuF} at 20~K and 1.75~K in zero field (ZF). As the temperature decreases, the spin relaxation increases due to the fluctuations of the magnetic Ni$^{2+}$ moments slowing down and entering the $\mu$SR time window \cite{J. H. Brewer, D. R. Noakes1}. At the lowest temperatures, the F-$\mu$-F signal is washed out by the magnetic field caused by Ni$^{2+}$ moments that relax the muon spin polarization. This is a similar process to the freezing of Ising-like Ho$^{3+}$ in LiHo$_{x}$Y$_{1-x}$F$_4$ \cite{J. Rodriguez}.

\begin{figure}[htbp]
\begin{center}
\setcounter{bottomnumber}{2}
\includegraphics[width=\columnwidth]{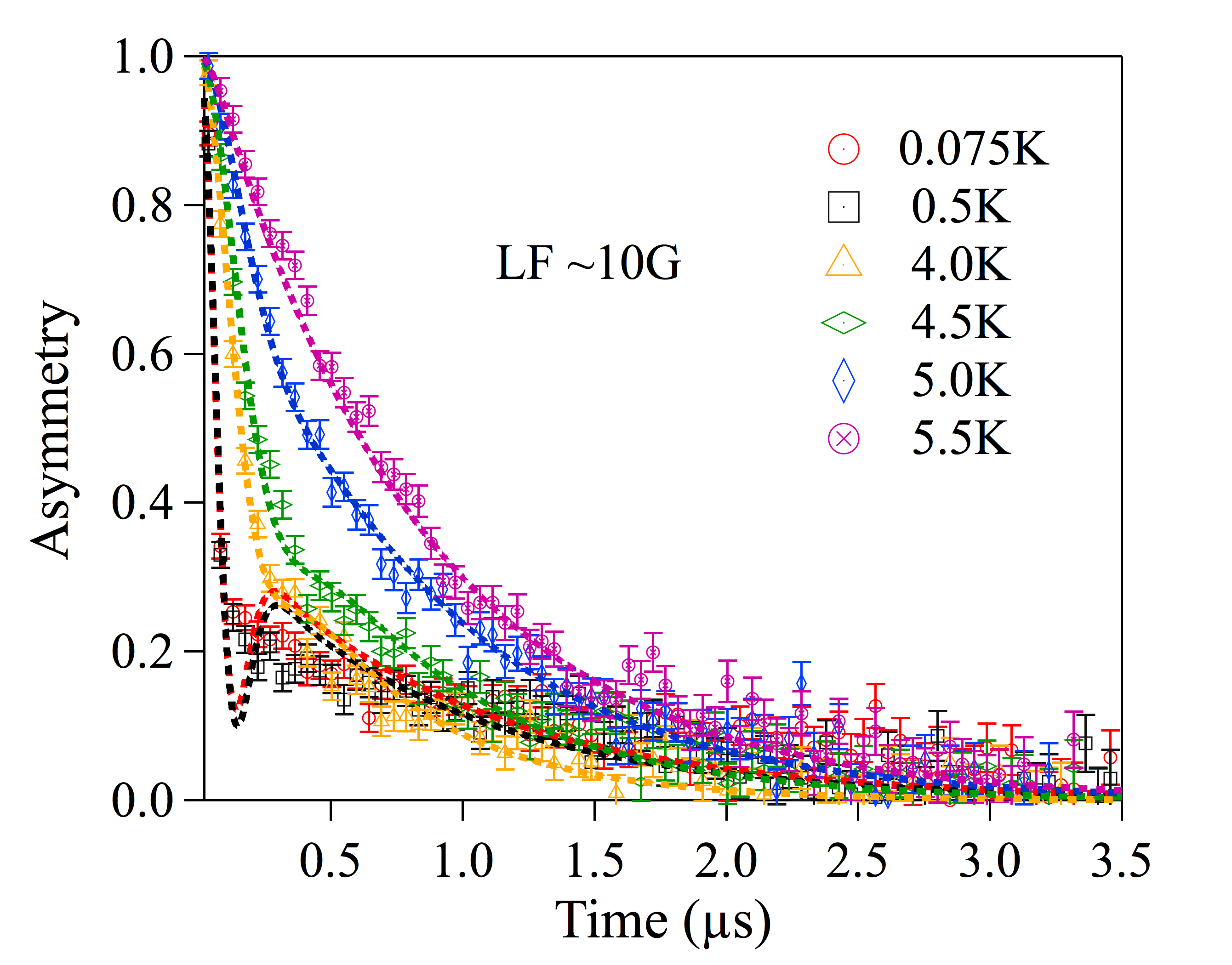}
\caption{NaCaNi$_2$F$_7$ $\mu$SR data measured at T= 0.075, 0.5, 4.0, 4.5, 5.0, and 5.5~K with a longitudinal field of 10~G. Dashed lines show the fits to the dynamical Gaussian Kubo-Toyabe function. \cite{Y. J. Uemura}}
\label{Fig: lineshapedata}
\end{center}
\end{figure}

Then we moved to dilution refrigerator to focus on the lower temperature range shown in Fig.~\ref{Fig: lineshapedata}, where the muons are dominated by fluctuating Ni moments rather than the F-$\mu$-F signal. As the temperature decreases from 6~K to 0.075~K, the change from a single relaxing component to two components with a rapid Gaussian relaxing front end and a slowly relaxing tail indicates the onset of freezing of the Ni moments and a relatively small spin fluctuation rate compared to the relaxation rate from the random distribution of the static local field. In a cubic system, a zero fluctuation rate $($static spins$)$ gives a non relaxing 1/3 tail corresponding to the 1/3 probability of the local field being parallel to the initial spin polarization, while a slow, but non-zero, fluctuation rate will relax this 1/3 tail \cite{Y. J. Uemura}.  

\begin{figure}[htbp]
\begin{center}
\setcounter{bottomnumber}{2}
\includegraphics[width=1.01\columnwidth]{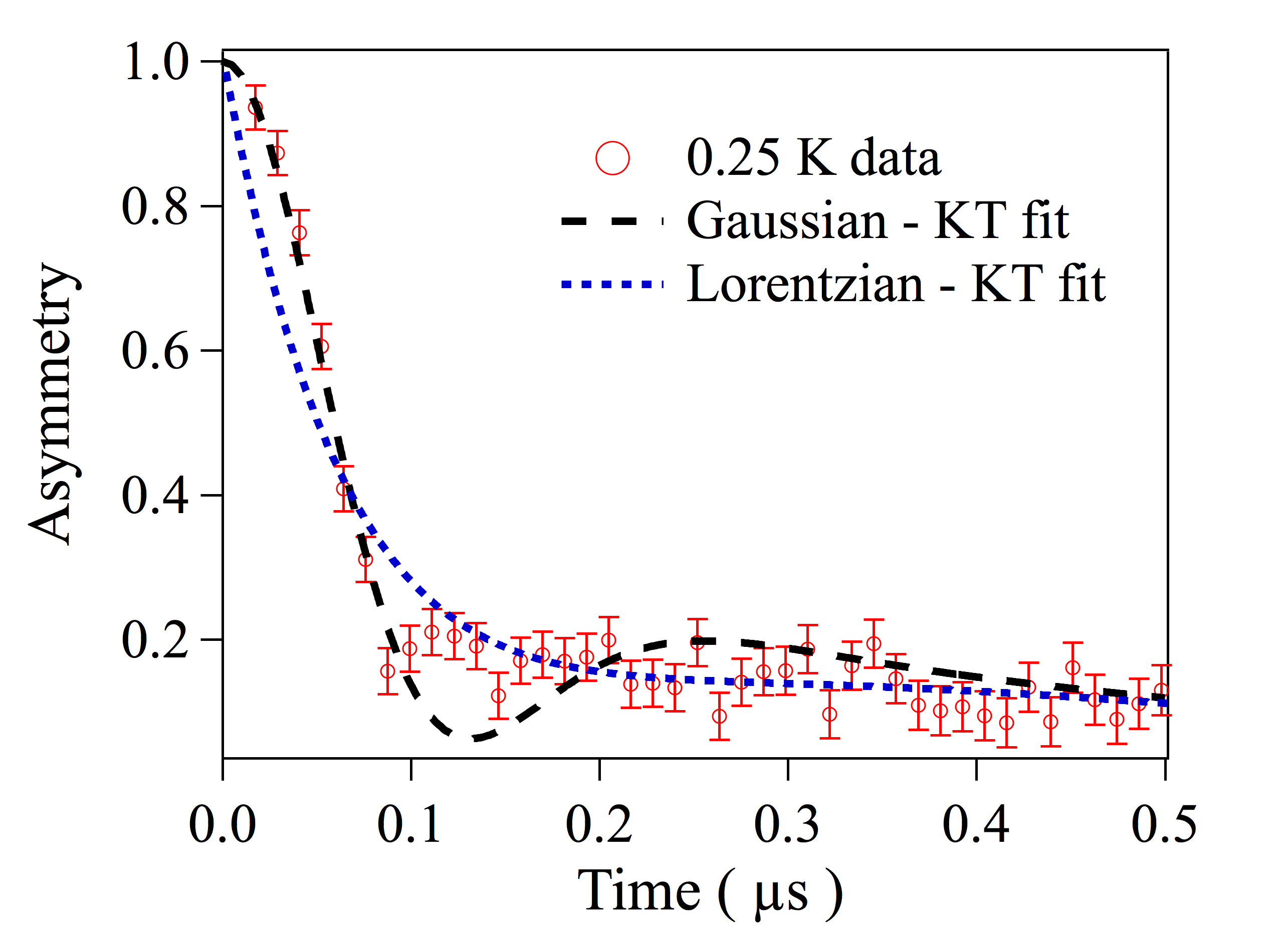}
\caption{First 0.5 $\mu$s of the NaCaNi$_2$F$_7$ $\mu$SR data measured at T= 0.25~K, showing the robust Gaussian feature of the fast front end. Dashed lines are the fitting results using a dynamical Gaussian Kubo-Toyabe function (black) and a dynamical Lorentzian Kubo-Toyabe function (blue).}
\label{Fig: Gaussian}
\end{center}
\end{figure}

These low temperature measurements were performed in a longitudinal field of 10~G. This small field serves to decouple any contribution to the asymmetry spectra from those muons stopping outside of the sample, leaving only contributions from the electronic magnetism in the sample. This data in Fig.~\ref{Fig: lineshapedata} shows a gradual increase of the static Gaussian relaxation rate with decreasing temperature. The fast relaxation is robustly Gaussian as shown in Fig.~\ref{Fig: Gaussian}, which is typical of dense static moments rather than dilute spins \cite{R. Kubo}. This indicates that the static field likely comes from the majority of the nickel moments, rather than arising from an effective dilute-system possibly caused by disorder. The data exhibit no sign of oscillations down to 0.075~K, which indicates an absence of long-range magnetic order.

\begin{figure}[htbp]
\setcounter{bottomnumber}{6}
\includegraphics[width=\columnwidth]{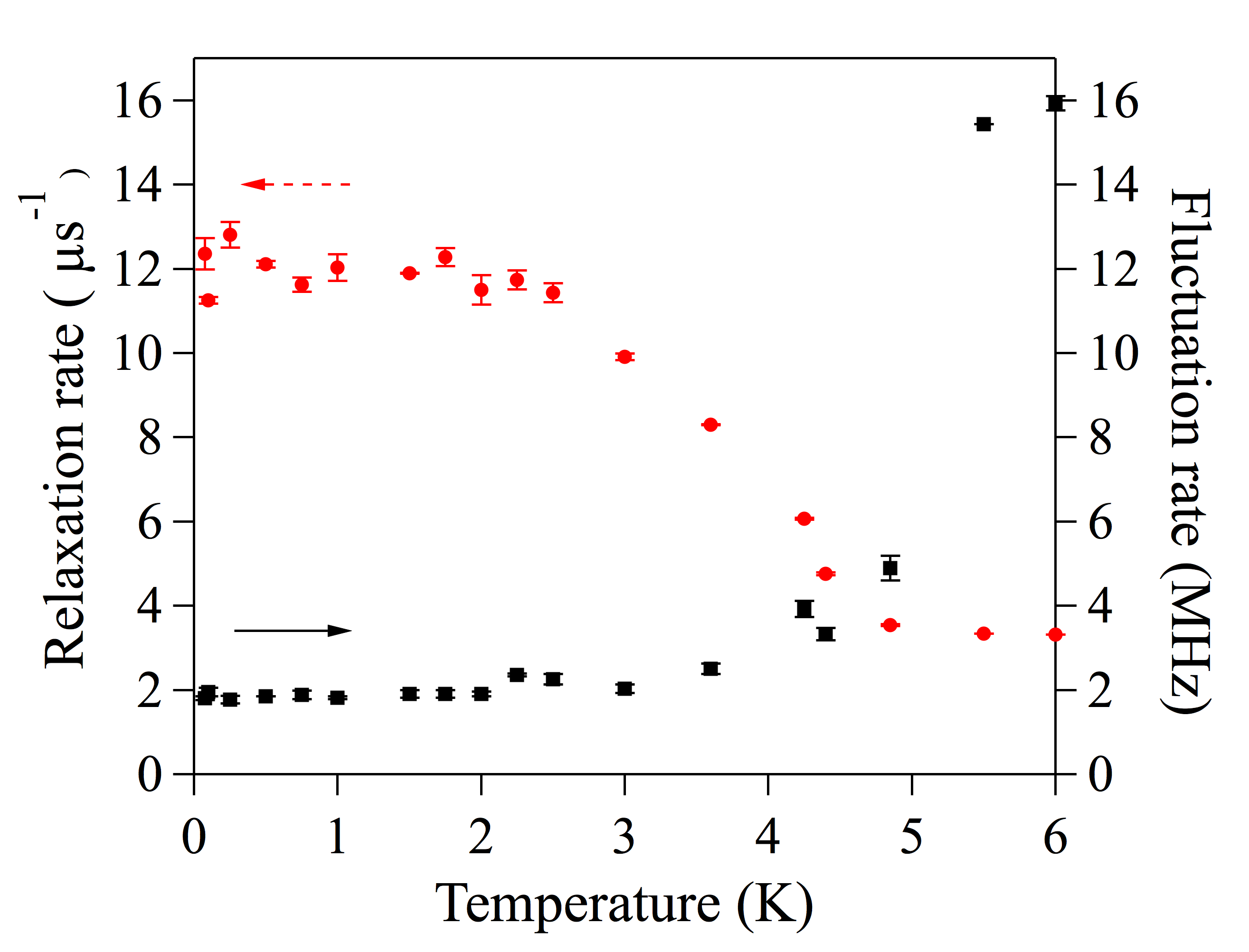}
\caption{ Fitting parameters of for the LF dynamic Gaussian Kubo-toyabe function used to fit the data in Fig. \ref{Fig: lineshapedata}. The relaxation rate (red circles) increases with decreasing temperature up a plateau at 12~$\mu{s}^{-1}$, while the fluctuation rate (black squares) decreases with decreasing temperature to a plateau at 2~MHz.}
\label{fig:relaxation rate}
\end{figure}

We fit the low temperature data with a dynamical Gaussian Kubo-Toyabe function (DGKTLF) \cite{Y. J. Uemura} and the results of this analysis are plotted in Fig.~\ref{fig:relaxation rate}. The relaxation rate reflects the width of the static internal field distribution and corresponds to an order parameter for the frozen spin state. The Gaussian relaxation rate also gives us an estimate of local field at the lowest temperature as ${\sigma}/\gamma_\mu = 140$~G ($\gamma_\mu$ being the gyromagnetic ratio of the muon). Comparing such a relatively small field strength to our simulation result described above of 2150~G, it seems that only a small fraction of the Ni moment freezes, leaving much of the spectral weight still fluctuating on the $\mu$SR timescale. The fluctuation rate decreases from 16~MHz at 6~K (above the transition) to 2~MHz below about 4~K and then remains roughly constant down to the lowest temperature (75~mK), which implies the presence of slow dynamics down to our lowest measured temperature in this system. This presence of spin dynamics in NaCaNi$_2$F$_7$ is also consistent with the existence of residual entropy as the magnetic entropy integrates to roughly R ln(2), which is significantly reduced from R ln(3) \cite{R. J. Cava}. Such persistent spin dynamics are frequently seen in a wide variety of geometrically frustrated systems although their physical origin is not understood. 

The minimum of the Kubo-Toyabe function does not agree perfectly with our data, which might mean that the true distribution is somewhat more complicated than a single Gaussian. However, we find that this single Gaussian Kubo-Toyabe nonetheless parameterizes the data well, and expect that there is little to be gained from attempting a more complicated fit. An example shown in Ref.\cite{D. R. Noakes3} for an increasing number of muon sites, each generating a standard static Gaussian distribution, demonstrates that the sum of the Kubo-Toyabe functions will smooth out the minimum. Here, since interactions between the muon and fluorines will distort the lattice, and Ca/Na site disorder will make muon sites inequivalent, we would expect a such distorted minimum.

\begin{figure}[htbp]
\includegraphics[width=\columnwidth]{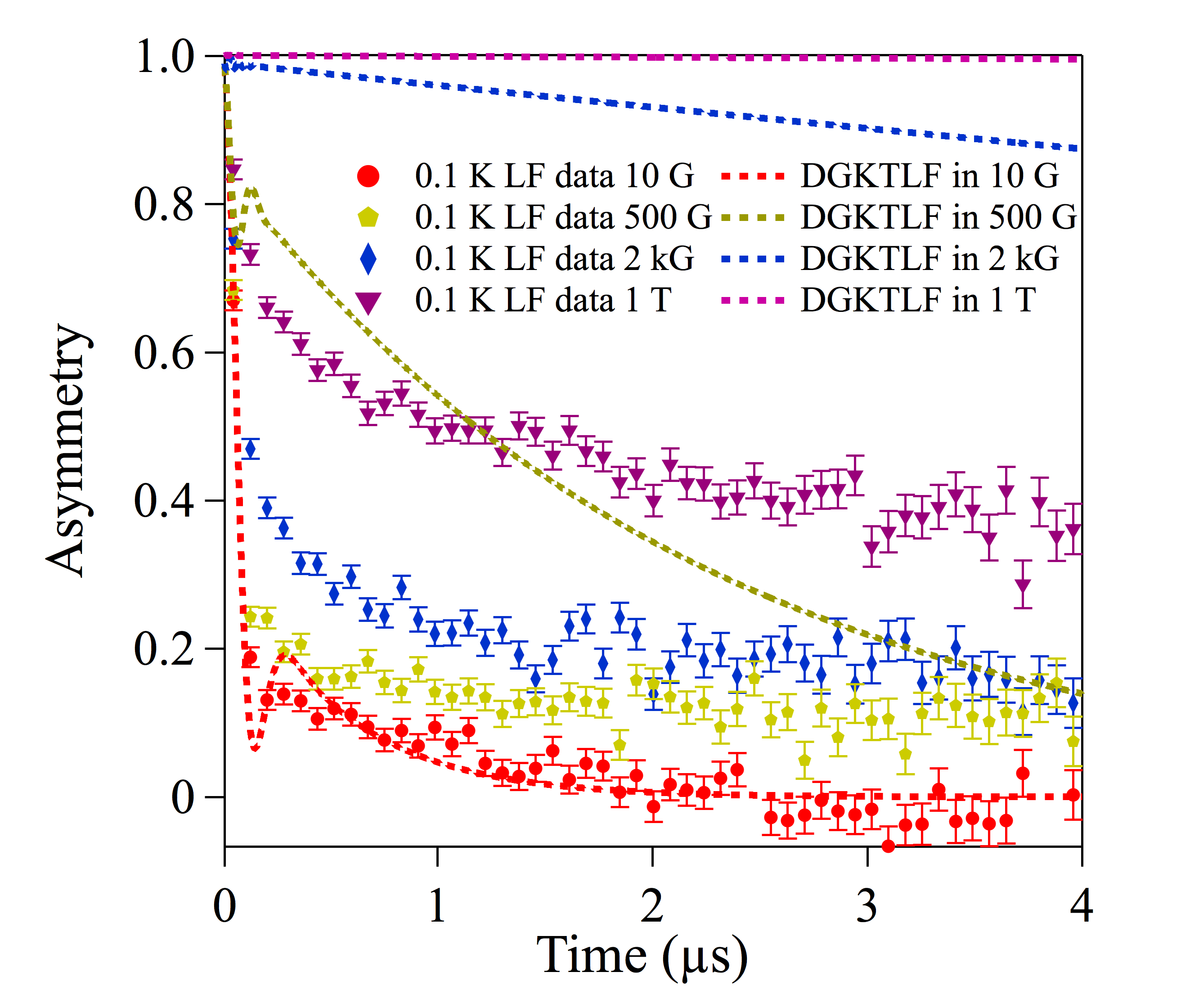}
\caption{$\mu$SR data at 0.1~K in selected fields up to 1~T, and LF Kubo-Toyabe function in the same selected field up to 1~T. These data demonstrate that the static relaxation is not fully decoupled even by large fields. The partially decoupled asymmetry spectra shows the coexistence of static and dynamic local fields acting on the muons.} 
\label{fig:LF lineshape}
\end{figure}

To better understand the dynamic/static nature of the local field, we performed longitudinal field (LF) $\mu$SR measurements. In a LF-$\mu$SR setup, the external field is applied in the direction of the initial muon spin polarization. In this case, the overall field seen by the muons is the vector sum of the internal field and the applied field. If the observed relaxation of the muon polarization comes from static magnetism, applying a longitudinal field will add this external field component to the internal field. This results in a net field seen by the muons that is closer to the direction of the muon spin, reducing the relaxation rate and relaxing asymmetry (called decoupling). In the strong collision model, a static signal will be nearly fully decoupled by an applied field that is a few times larger than the internal field measured in zero field, and we can fit this using a longitudinal field Kubo-Toyabe function. However, if the relaxation of the ZF $\mu$SR signal comes from dynamics, the signal will not be decoupled by an applied field of this magnitude, as there is no well-defined weak internal field. In this case, the signal will only slowly decouple, and relaxation will continue to be apparent up to relatively large applied longitudinal fields \cite{Uemura3}.

Our LF results shown in Fig.~\ref{fig:LF lineshape} are not well described by the DGKTLF: we only see partial decoupling of the fast relaxation in contrast to expectations based on the ZF analysis. This indicates that a substantial portion of the relaxation comes from spin dynamics, which is inconsistent with the 2-component form of the ZF relaxation which implies slow dynamics. A possible resolution of this apparent paradox could be that the strong collision model might not be applicable in this situation as it assumes a field independent fluctuation (collision) rate. Our observations imply that the application of a LF enhances the spin fluctuations, which would also end up with a non-decoupled signal. This persistent Gaussian relaxation associated with dynamics brings to mind the Kagome material SrCr$_{x}$Ga$_{12-x}$O$_{19}$ (SCGO), the pyrochlores Yb$_{2}$\textit{B}$_{2}$O$_{7}$ (B = Ti,Ge), and other frustrated systems, where similar relaxations were observed that were not decoupled by appropriate longitudinal fields \cite{Uemura2, Ortenzio, Hallas2}.

In conclusion,  we have presented a $\mu$SR study of spin dynamics in NaCaNi$_2$F$_7$. We found a partial spin freezing transition at 4~K to a spin-glass like state involving the majority of the Ni moments. We also observed a residual temperature independent spin fluctuation rate down to 0.075~K, which shows that there are persistent spin fluctuations at low temperature, as is common in highly frustrated systems. As this system combines intense geometric frustration with chemical disorder caused by randomly distributed Ca and Na ions on the \textit{A}-site, the underlying cause of the dynamics is not yet fully understood. Our data, along with previous work showing existence of residual entropy and similar results in the titanium pyrochlores, suggest that bond disorder in frustrated systems results in a glassy ground state \cite{R. J. Cava}. This similarity offers a great opportunity to explore the related properties of the new transition metal fluoride pyrochlores and the titanium oxide pyrochlore family. Finally, our DFT calculations and simulations suggest the muon position in NaCaNi$_2$F$_7$ is somewhat off centre between two inequivalent fluorine ions. The calculated field at the simulated muon site is 15 times larger than we observe experimentally, providing evidence that the full Ni moment is not frozen, and that much of the fluctuations remain above the $\mu$SR time window. Further neutron scattering studies on the magnetic excitations are needed to explore the magnetic anisotropy, which will also be helpful for understanding the universal persistent dynamics in frustrated magnets.

We greatly appreciate useful discussions with Wen Huang. We also greatly appreciate the support of personnel at TRIUMF during the $\mu$SR measurements. Work at Princeton was supported by the US Department of Energy, Division of Basic Energy Sciences, through the Institute for Quantum Matter, grant DE-FG02-08ER46544. Work of the Columbia group has been supported by NSF DMR-1436095 (DMREF) and NSF DMR-1610633, and Reimei Project of Japan Atomic Energy Agency. The DFT calculations were performed using computational resources of the Thunder Bay Regional Research Institute, Lakehead University, and Compute Canada (Calcul Quebec). 

Work at McMaster ( G.~M.~Luke, O.~Rubel ) was supported by the Natural Sciences and Engineering Research of Council of Canada.

\end{document}